# Aquaporin-1 can work as a Maxwell's Demon in the Body

Liangsuo Shu[1]*, Yingjie Li*[2], Xiaokang, Liu[1] Xin Qian[1], Suyi Huang[1], Shiping Jin[1], and Baoxue Yang[2]

[1]Innovation Institute, School of Energy and Power Engineering, Huazhong University of Science & Technology. 1037 Luoyu Road, Wuhan, China;

[2]State Key Laboratory of Natural and Biomimetic Drugs, Department of Pharmacology, School of Basic Medical Sciences, Peking University, Beijing, China;

*contributed to this work equally

Xin Qian is at Mechanical Engineering, University of Colorado Boulder now.

Address correspondence to:

**Shiping Jin**

Innovation Institute, School of Energy and Power Engineering,

Huazhong University of Science & Technology.

1037 Luoyu Road, Wuhan, 430074, China;

E-mail: jinshiping@hust.edu.cn

Tel: 0086-027- 87542618

**Baoxue Yang**

Department of Pharmacology, School of Basic Medical Sciences,
Peking University,
38 Xueyuan Road, Haidian District,
Beijing, 100191, China.
E-mail: baoxue@bjmu.edu.cn
Fax: 0086-010-82805622

Key words: Maxwell's Demon, Aquaporin-1, Nano-osmosis

**Abstract**

Aquaporin-1 (AQP1) is a membrane protein which is selectively permeable to water. Due to its dumbbell shape, AQP1 can sense the size information of solute molecules in osmosis. At the cost of consuming this information, AQP1 can move water against its chemical potential gradient: it is able to work as one kind of Maxwell's Demon. This effect was detected quantitatively by measuring the water osmosis of mice red blood cells. This ability may protect the red blood cells from the eryptosis elicited by osmotic shock when they move in the kidney, where a large gradient of urea is required for the urine concentrating mechanism. This finding anticipates a new beginning of inquiries into the complicated relationships among mass, energy and information in bio-systems.

Since 1867, Maxwell's Demon has been one of the focuses of thermodynamic debates and a hot topic of popular science[1]. After a series of theoretical works by many researchers, the demon had been proved to be able to convert information into free energy without contradicting the second law of thermodynamics[2]. In fact, some "demons" have been created in different ingenious experiments[3–5] and more theoretical models to implement Maxwell's Demon have been proposed[6–11]. However, it is necessary to determine whether any Maxwell's Demons, which can convert information into free energy, exist in natural world and their scientific significance.

Osmosis plays an important role in areas such as regulating water balance across cellular membranes[12], the immune responses to pathogens of lymphocyte perforin[13] and the membrane attack complex/perforin [14], and the function of pore-forming toxins[15]. In thermodynamics, it is generally accepted that osmotic pressure is the result of the chemical potential difference of the solvent across a semipermeable membrane, but the debate in osmosis dynamics never ends[16]. The discovery[17] and follow-up studies[18,19] of a nature nanoscale channel, aquaporin (AQP) give us a good opportunity for a better understanding of osmosis[20].

The researches of AQP osmosis reveal an unexpected phenomenon: the reflection coefficients of small impermeable solutes to AQP1 are smaller than 1 and have a close relation with their molecular sizes [21–23](*Solomon-Hill effect*). In a previous work, we got the quantitative explanation of this effect using an analytical method based on molecular dynamics[20]. The dumbbell shape of AQP1 makes it able to sense the size information of solute molecules in osmosis. However, this ability enables AQP1 to move water molecules against their chemical potential gradient in some cases. Considering the information consumption during the osmosis in these cases, we find Aquaporin-1 can work as a Maxwell's Demon which can convert information into free energy.

The reflection coefficients of impermeable solutes (urea, mannitol, glucose) to AQP1 were measured using red blood cells(RBCs) lacking urea transporter-B (UT-B), from UT-B knockout mice to rule out the possible influences of the permeability of urea through UT-B[24,25].

**Results**

**reflection coefficients of impermeable solutes**

The osmotic pressure of a dilute solution is usually described by the famous Van't Hoff equation

$$\pi = \sigma cRT \tag{1}$$

where $\pi$ is the osmotic pressure, $c$ is the molar concentration of the solute, $R$ is the molar gas constant, $T$ is the thermodynamic temperature, and $\sigma$ is the reflection coefficient. $\sigma$, introduced as a phenomenological coefficient[26], was defined as the fraction of a certain solute that does not permeate the membrane, its retention. This definition has resulted in an inference that $\sigma$ of a completely impermeable solute must be 1. However, many experiments about AQP1 clearly indicate that $\sigma$ values of some impermeable solutes are smaller than 1 and have a close relation to their molecular size[21–23], the *Solomon-Hill effect*. It was found that this effect is a result of the special dumbbell-like shape of AQP1[20,22]. As a remarked characteristic of all AQPs, the dumbbell-like shape is consists of an extracellular vestibule, a cytoplasmic vestibule, and a narrow channel connecting them [18]. The vestibule surrounded by loops allow the incursions of solute molecules to contribute part of momentums in varying degrees depended on their molecular sizes[20,22].

In the channel of AQP, there is usually at least one filter which determines its selectivity. For AQP1, the filter is too narrow to allow any other molecules beside water to pass through. AQP1 contributes the main water permeability of RBCs and makes it an ideal materiel to measure the osmosis. In the experiments of Toon and Solomon[23], they used common RBCs, which also expressed UT-B, as the materials to measure the reflection coefficient of solutes. This can lead to some degree of errors. On one hand, besides AQP1, UT-B and its plasma membrane also make some contribution of the water permeability of RBCs[24,25,27]. On the other hand, little solutes, such as urea, can permeate through UT-B and reduce the measured value of reflection coefficients. Therefore, Solomon-Hill effect was also suspected to a confounding effect of rapid diffusional urea transport in some literatures [24].

To rule out the influence UT-B, we used UT-B null (UT-B-/-) RBCs to measure the reflection coefficients of solutes for AQP1 with stopped-flow. $\sigma$s of solutes are calculated by comparing the initial rates of red cell volume changes in different solutions[23,28,29]. The dead times of our experiments were about 200 millisecond. Therefore, the slopes from 200 millisecond to 300 millisecond were used. For UT-B null RBCs (urea can be regarded as one impermeable solute), a widely used method of exponential function fitting [25,30–32] in reconstituted aquaporin osmotic permeability was also applied to calculate the reflection coefficients. The results calculated from two methods agreed well with each other.

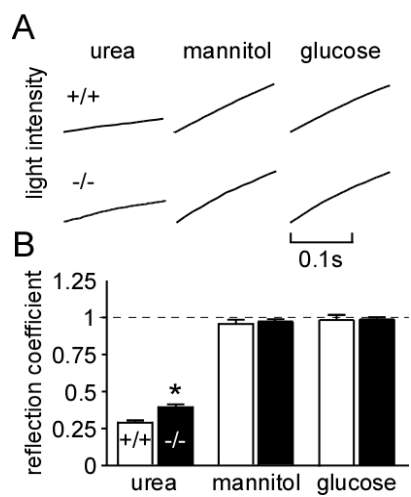

**Fig. 1. Water permeability and reflection coefficients of AQP1 with different solutes in RBCs.** Osmotic water permeability was measured from the time dependence of erythrocyte volume, in response to a 125-mM inwardly directed solute gradient, by stopped-flow light scattering. A, Representative curves showing water permeability in RBCs from wild-type mice (+/+) and UT-B null mice (-/-) measured at 37 ℃ under urea, mannitol or glucose and using a 125 mM gradient. B, averaged reflection coefficients of AQP1 with different solutes for experiments conducted as in A (mean ± S.E. n≥3 )

Our experimental results showed that the reflection coefficients of solutes for AQP1 increase with their molecular sizes (Fig.1). It was found that rapid diffusional urea transport through UT-B does make a difference to the reflection coefficients of urea [28], while no remarked difference was found for mannitol and glucose. However, this influence is rather limited compared with the size effect. For RBCs of UT-B null mice, $\sigma_{urea}$ is reduced to 0.39 from 1 because of the size effect (the reflection coefficients of glucose is regarded as 1, $\sigma_{glucose}=1$). For erythrocytes of wild mice, $\sigma_{urea}$ only reduces 0.09 to 0.30 because of the urea transport through UT-B. This can be attributed to the high water permeability of AQP. Both water osmosis through AQP1 and urea transporting through UT-B are non-equilibrium processes, during which time is an important factor. From the works of Zhao[33], it can be found that the osmotically induced water efflux of red blood cells is much faster than urea transporting through UT-B.

Therefore, the difference in response times of the two processes isolates them to a great extent. By ruling out the influence of UT-B, our improved experiments provided more accurate measurements of *Solomon-Hill* effect.

**Thermodynamics of Maxwell's Demon in Our Body**

The small reflection coefficients of impermeable solutes have also brought us a new question. Suppose the simplest of conditions: there is an erythrocyte and the solutes for the intracellular and extracellular solutions are pure glucose and pure urea respectively. When $\sigma_{urea}c_{urea} < c_{glucose} < c_{urea}$, water is moved into the red blood cell through AQP1 against its chemical potential gradient. It is an entropy decreasing process which seems to break the second law of thermodynamics (Fig.2*a*). The entropy decrease($ds_{1\text{-}2}$) can be calculated as below,

$$ds_{1-2} = \frac{(\mu_{ex} - \mu_{in})dn}{T} \approx R(c_{glucose} - c_{urea})dV < 0 \qquad (2)$$

where $\mu_{ex}$ and $\mu_{in}$ are the chemical potential of extracellular and intracellular water respectively, $dn$ and $dV$ are the transferring amount and volume of water through AQP1 respectively.

However, the real osmotic pressure difference sensed by the AQP1 in its osmosis is

$$\Delta\pi = RT(c_{glucose} - c_{urea}\sigma_{urea}) \qquad (3)$$

and the entropy change during osmosis from the view of AQP1 is

$$ds_{1'-2'} = \frac{\Delta\pi dV}{T} = R(c_{glucose} - \sigma_{urea}c_{urea})dV > 0 \qquad (4)$$

Therefore, for AQP1, the osmosis is a spontaneous process of entropy increment, during which the second law of thermodynamics is strictly followed(Fig.2b).

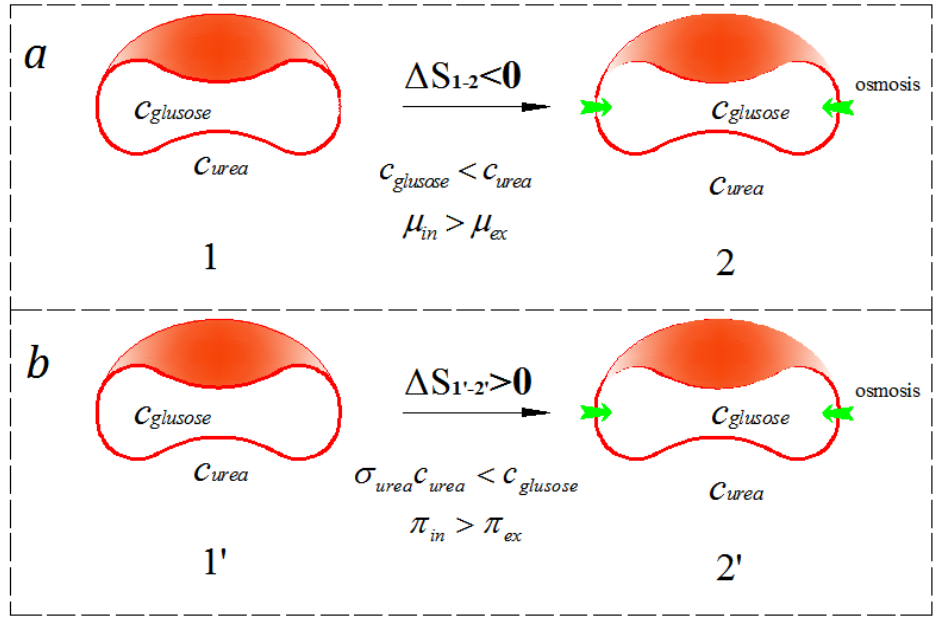

Fig. 2 **Entropy changes of a same osmosis process from the views of different observers.** The green arrowhead is the water transferred into the red cell through AQP1. *a*, from the view of a common observer, it is an entropy decreasing process during which the water is moved against its chemical potential gradient; from the view of AQP1 which can sense the information of molecular size, it is a spontaneous process of entropy increment driven by osmotic pressure.

Comparing Equation (2) and (4), it can be found that the entropy functions of a same osmosis process are different for different observers. What's puzzling is that the second law of thermodynamics is broken for an observer while is strictly followed for another. There must be a missing entropy hidden in this puzzling appearance. The molecular size is one kind of information of solute. The special dumbbell shape of AQP1 makes it have the ability to "read" this information, which appears as the reflection coefficients of solutes in equation (4). This ability generates an *information gap* between AQP1 and a common observer without this ability. During the osmosis, this information was continuously consumed. The Landauer's principle linking information and thermodynamics had been confirmed by the experiments of Bérut et al[5]. By adding the entropy increase caused by the consumption of information, the whole entropy changes from the view of an observer from conventional thermodynamic will also be greater than 0. Taken together, AQP1 in fact can work as one kind of Maxwell's Demon in the body: it can move water against its chemical potential gradient at the cost of consuming information (as shown in Fig.3).

The information consumed during the osmosis can be defined as follows,

$$di = ds_{1'-2'} - ds_{1-2} = Rc_{urea}(\sigma_{urea} - 1)dV < 0 \qquad (5)$$

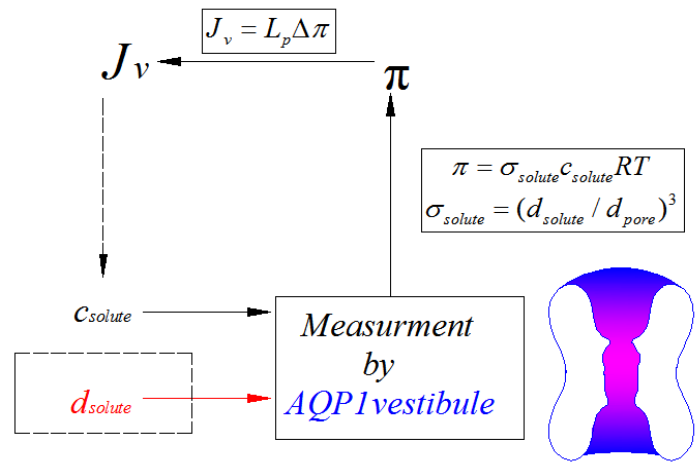

**Fig.3. Feedback loop of AQP1 when working as a Maxwell's Demon.** $d_{solute}$ and $d_{pore}$ are the effective diameters of solute molecule and AQP1 vestibule respectively, $c_{solute}$ is solute concentration. $J_v$ is the osmotic water flux driven by the osmotic pressure

difference($\Delta\pi$) and $L_p$ is hydraulic conductivity. The equation describing the relationship between reflection coefficient($\sigma_{solute}$) and relative size of a solute molecule to the pore comes from reference[20]. In the bottom right corner, there is a diagram for the dumbbell-like shape of AQP1: an extracellular vestibule (blue), a cytoplasmic vestibule (blue), and a narrow channel(pink) connecting them. The measurement of solutes molecular size and concentration by the vestibule of AQP1, the information translation form molecular size to reflection coefficient which (together with solute concentration) determines the osmotic pressure, and the control of osmotic water flux according to osmotic pressure compose a complete feedback loop[34] which generates information flow. Comparing with a common thermodynamics observer which can only get the information of solute concentration, AQP1 can work as a Maxwell's Demon in same cases (such as the situation in Fig.2). During the osmosis, the information of the memory of Maxwell's demon is continuously consumed.

When $\sigma_{urea}c_{urea} = c_{glucos}$, a thermodynamics balance from the view of AQP1 will be reached. Integrating the above equation from initial state to this balance, we can get the whole information of AQP1 about the molecular size of solutes ($I$)

$$I = Rc_{urea}V_0(1-\sigma_{urea})(\frac{c_{glucose}}{\sigma_{urea}c_{urea}}-1) \tag{6}$$

where $V_0$ is the initial cell volume.

For a normal red blood cell with UT-B, not all information can be used by AQP1 in its osmosis: a part of information (about 23%) loses because of the diffusional urea transport through UT-B during the osmosis.

**Mutual transformation of information and energy in aquaporin osmosis**

The information of size effect is a result of the special structure of AQP1, which is controlled by the genetic information in DNA. During the expression of this information, some energy ($E_1$) is consumed. This special structure of AQP1 makes it has the ability to read the information of molecular size of solute. Then, during the osmosis of red blood cells in the kidney with a large gradient of urea, AQP1 can move water against its chemical potential gradient at the cost of consuming the information of size effect: information is converted back into free energy again ($E_2$). In the whole process, information acts as one kind of energy storage medium.

$$I_{DNA} + E_1 \rightarrow I_{AQP1} \rightarrow E_2 \qquad (7)$$

Some kinds of information engines, which can convert information into free energy, have been discussed in theory[6,8]. From the above discussion, an artificial information engine may be realized by a bionics design from the idea of AQP1 osmosis in the kidney.

**Discussion**

The way how AQP1 works as a Maxwell's demon confirms the suspicion of James Maxwell about one and a half century ago. However, the special structure of AQP1 is a result of a long period of biological evolution for a better adaptability to the changing environment, knowing nothing about Maxwell's theoretical prediction. Therefore, is there any possible significance of this Maxwell's demon in our body?

In the kidney, there is a large urea gradient which is required for the urine concentrating mechanism[35]. The osmotic water of red blood cells caused by urea can lead to the volume change of red blood cells when they move from the renal cortex to the renal medulla, or vice versa. A smaller reflection coefficient of urea ($\sigma_{urea}$) for AQP1 means a smaller osmotic water, and then a smaller volume change. This can reduce the eryptosis elicited by osmotic shock. Therefore, a small $\sigma_{urea}$ for AQP1 is necessary to protect the red blood cells from the eryptosis elicited by osmotic shock. In our experiments, for red blood cells of UT-B null mice, the whole volume change caused by urea gradient is only about half of that caused by glucose gradient at a same concentration; for red blood cells of wild mice, the upper limit of this value is 0.61.

What's more, the osmotic water of red blood cells caused by urea can also weaken the urea gradient between internal and external the red blood cells and slow down the urea transferring through UT-B. If the osmotically induced water efflux weakens when urea has a small reflection coefficient, this weakening effect of urea transfer can be eased to some extent. Therefore, a small $\sigma_{urea}$ to AQP1 should make a beneficial contribution to the fast urea absorbing and releasing across the erythrocyte plasma membrane through UT-B, which plays an important role in the urine concentrating mechanism[36].

**Materials and Methods**

**Transgenic mice**

UT-B knockout mice were generated by targeted gene disruption as previously reported[37]. Wild-type and UT-B null mice at 8~10 weeks old were housed at constant room temperature (23±1°C) and relative humidity (50%) under a regular light/dark schedule (light on from 7:00A.M. to 7:00P.M.) with free accessing food and water before experiments.

**Reflection coefficients measurement using stopped-flow**

Fresh red blood cells obtained by tail bleeding (100~200 μl/bleed) were washed three times in phosphate-buffered saline (PBS) to remove serum and the cellular buffy coat. The composition of PBS is (in mM): NaCl, 154; $Na_2HPO_4$, 10; $NaH_2PO_4$, 10. Stopped-flow measurements were carried out by stopped flow spectrometer SX20 (Applied Photophysics, UK) on a Hi-Tech Sf-51 instrument. For measurement of reflection coefficients, suspensions of red blood cells (~0.5% hematocrit) in phosphate buffered saline were subjected to a 125 mM inwardly directed gradient of urea, mannitol and glucose. The kinetics of decreasing cell volume were measured from the time course of 90° scattered light intensity at 530 nm wavelength [38].

**Computation of reflection coefficients**

$\sigma$s of solutes are calculated by comparing the initial rates of the red blood cell volume change at *zero-time*[39] in different solutions

$$\frac{dV}{dt}(t=0) = P_f v_w \frac{S_0}{V_0} \sigma c RT \qquad (8)$$

Where $V$ is the cell volume normalized, $v_w$ is the partial molar volume of water, $S_0/V_0$ is the initial cell surface-to-volume ratio, $c$ is the initial concentration of solute which causes the osmotic challenge[23,28,29]. For every experiment, the measurement was repeated 10 or 20 times. The *zero-time* was determined when most scattered light intensities are linear varying and parallel to each other. The dead time was mainly caused by the mixing process of solution and a red blood cell suspension (some new approach has been making to reduce it recently[28]). A part of signals which deviated significantly from the main tendency were neglected and the average value of the remaining signals was regarded as the result of the experiment. There were three measurements for every solute.

For UT-B null red blood cells, a widely used method [25,30–32] developed by van Heeswijk and van Os 1986 [40] was also used to calculate reflection coefficients.

$$y = \alpha e^{-kt} + \beta \qquad (9)$$

where $y$ is the scattered light intensity normalized, $\alpha$, $k$, $c$ are three parameters. When comparing the water permeability of different channels or membranes to the same osmotic challenge (both the solute and its concentration are the same), $k$ was direct proportional to water permeability ($P_f$). However, when calculating the $\sigma$s of different solutes to the same channel (such as AQP1), we needed to use $\alpha$.

$$\begin{cases} \alpha = \dfrac{V_0 - b}{AV_0}(1 - \dfrac{c_{i0}}{c_{o0}}) \\ c_{o0} = c_{i0} + \sigma c \end{cases} \qquad (10)$$

where *A* is a system parameter, which can be regarded as a constant here; $V_0$ is the initial cell volume, *b* is the osmotically inactive volume, $c_{i0}$ is the initial cell osmolarity and $c_{o0}$ is the initial equivalent outside osmolarity, *c* is the concentration of solute (urea, mannitol or glucose) and $\sigma$ is its reflection coefficient. The detailed derivation of above equation will be given in support information.

**Computation of whole volume change**

Assuming the change of the amount of solutes inside the cell can be ignored when osmotic equation is reached[40], we can get

$$(V_\infty - b)C_{o0} = (V_0 - b)C_{i0} \tag{12}$$

$$V_\infty - V_0 = (V_0 - b)(1 - \frac{C_{i0}}{C_{O0}}) = \alpha A V_0 \tag{13}$$

From above equation, we can discover that the whole volume change is in direct proportion to *α*.

$$\frac{(V_\infty - V_0)_{urea}}{(V_\infty - V_0)_{glucose}} = \frac{\alpha_{urea}}{\alpha_{glucose}} \tag{14}$$

However, UT-B is an efficient urea channel[24]. The urea transfer of wild mice red blood cells during the osmosis can't be ignored completely and the whole volume change of red blood cells is reduced to a certain extent. Therefore, for wild mice red blood cells with UT-B, the actual whole volume change is smaller than the value calculated by above equation.

**Acknowledgements**

The authors thank A.E Hill (Physiological Laboratory, University of Cambridge), Peter Agre (JHMRI, Johns Hopkins University), and John Mathai (BIDMC, Harvard Medical School) for useful discussions and valuable suggestions through e-mail. The helps in language or illustration from Miss Hu Qiyi, Miss Zhang Yibo and Mr. Oliver Robshaw are also gratefully acknowledged.